\documentclass[sigconf,screen,nonacm] {acmart}
\AtBeginDocument{%
  }

\setcopyright{cc}
\setcctype[4.0]{by}

\begin{document}

\title{Creative Task Cards for Reflection, Self-Efficiency and Self-Regulation in CS1 Introductory Programming: Initial Insights}

\author{Corey Ford}
\orcid{0000-0002-6895-2441}
\affiliation{%
  \institution{University of the Arts London}
  \city{London}
  \country{UK}
}
\email{c.ford@arts.ac.uk}

\author{Yinmiao Li}
\orcid{0000-0001-9570-3961}
\affiliation{%
  \institution{Northwestern University}
  \city{Evanston}
  \country{USA}
}
\email{yinmiaoli@u.northwestern.edu}

\author{Rosa Van Koningsbruggen}
\orcid{0000-0002-1652-2888}
\affiliation{%
  \institution{Bauhaus-Universit{\"a}t Weimar}
  \city{Weimar}
  \country{Germany}
}
\email{rosa.donna.van.koningsbruggen@uni-weimar.de}

% \renewcommand{\shortauthors}{Ford et al.}

% \textbf{REVIEWS:}
% [TODO] The task cards section could be improved by linking the design more to previous work on reflection. \newline

% Abstract
\begin{abstract}
Computer Science students in introductory programming courses struggle with emotional challenges such as low levels of sustained interest and a lack of self-belief. However, students are typically introduced to disciplinary strategies (such as debugging strategies) and not the metacognitive nor self-regulation strategies that could help them overcome such emotional challenges. This paper addresses this issue by introducing creative task cards designed to support students' reflection, self-efficacy, and self-regulation in programming assignments. To evaluate the cards, twenty-nine students used them to complete a mock coding assignment in-class and participated in surveys and focus groups. Our initial qualitative insights suggest that students: found value in taking breaks, particularly when they could leave the classroom to reflect with peers; could use drawing to better reflect on feelings of cognitive overload; and felt more capable progressing with their work once breaking it into smaller steps.
\end{abstract}

\begin{teaserfigure}
 \centering
  \includegraphics[width=.98\textwidth]{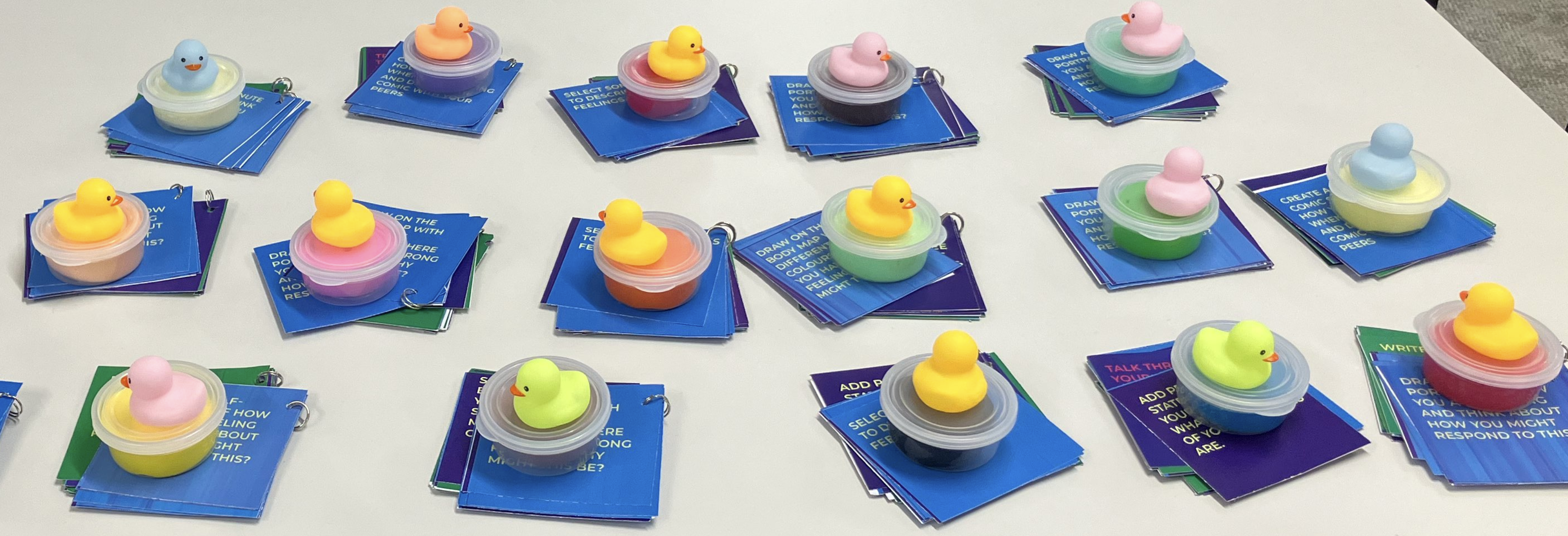}
  \caption{The creative card kits.}
  \label{fig:teaser}
\end{teaserfigure}

% % CCSXML


% %Keywords
% \keywords{Do, Not, Use, This, Code, Put, the, Correct, Terms, for,


\maketitle

\textbf{Reference:}\newline
Corey Ford, Yinmiao Li and Rosa Van Koningsbruggen. 2026. Creative Task Cards for Reflection, Self-Efficiency and
Self-Regulation in CS1 Introductory Programming: Initial Insights. In \textit{Proceedings of The First Reflection in Creative Experience (RiCE) Workshop (RiCE W1)}. ACM Creativity \& Cognition 2026, London, UK.

%=========================================================

\section{Introduction}\label{sec:introduction}
Computer Science (CS) students in introductory courses struggle to learn programming, with prominent issues including low levels of persistence and a lack of sustained interest \cite{Porter2013Retaining}. The lack of persistence and interest is interwoven with students' emotions and metacognitive judgements of self. For example, students often have low self-belief in their computational thinking skills \cite{Li2024} and draw comparisons to their peers which they perceive to have stronger CS skills \cite{Li2026}. Indeed, whilst students assess themselves based on their self-expectations and beliefs about others’, this contrasts with other's actual skill levels; students assess themselves harshly. Chen et al. \cite{Chen2024Understanding} suggested that ``interactions with professionals, instructors, course policy, and peers can help students understand what experiences are expected while learning to program, while the absence of these interactions may lead to negative self-assessments due to a lack of accurate information about what to expect.''

The negative affective and metacognitive experiences described above can lead to negative spirals, cyclically derailing the problem-solving process and students' persistence. We suggest that students thus need support in both developing disciplinary skills and in navigating the affective and metacognitive dimensions of their learning experiences. 

It has been demonstrated that students’ negative affective and metacognitive experiences are influenced by their knowledge of strategies to overcome them \cite{Medeiros2019Systematic}. Students do not know what, when and how to use both disciplinary (e.g. debugging strategies in CS) and regulation strategies (e.g. taking breaks). Students who are equipped with suitable strategies stay engaged while encountering struggles, and the effective use of these strategies can prevent the spiral of negative emotions and unhelpful self-assessment \cite{Garlandetal2010}.  %being able to take breaks and socialise with peers to understand that negative feelings whilst learning programming is normal.

However, in traditional CS1 courses, students often do not have the space to talk about their emotions, how they subjectively feel or evaluate themselves, their beliefs of themselves, and so forth. Whilst existing research in the learning sciences has explored cognitive strategies to support students, there is opportunity to incorporate space for students to reflect on their emotions and self-belief in their programming experiences, then share and discuss with their classmates, to propose strategies that could help them overcome challenges. 

We thus developed a card kit with small tasks that explore fun ways to encourage self-efficacy and self-regulation in CS programming and tested this in the classroom. \textbf{Our research question is: }
\begin{itemize}
    \item What if we ask students to use creative task cards in peer-to-peer discussions which prompt reflection on emotional aspects of programming. How does this support students’ self-efficacy and self-regulation?
\end{itemize}

Twenty-nine students used the creative task cards in-class whilst completing a mock coding assignment and participated in both surveys and focus groups. Our preliminary findings illustrate student's opinions and perceptions of various self-regulation and disciplinary strategies. Our main contributions are the creative tasks card kit and descriptions of how students recognised their own emotions, and selected strategies to overcome this.

\begin{figure*}[ht]
    \centering
    \includegraphics[width=\linewidth]{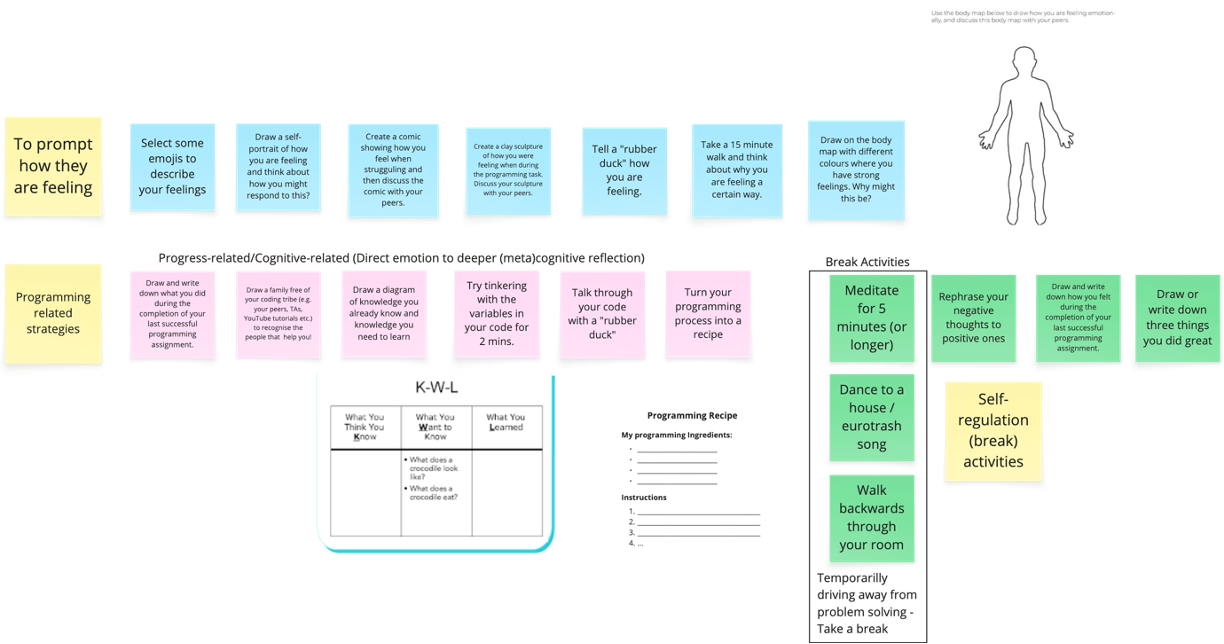}
    \caption{Brainstorming for the creative task cards}
    \label{fig:brainstorm}
\end{figure*}

\section{Creative Task Cards for Self-Efficiency and Self-Regulation} \label{sec:prototype}
To develop our cards, we examined literature from both the learning sciences and the arts. We set the following desired outcomes to assess the usefulness of the literature for our card deck: 
\begin{itemize}
   \item  Students are able to have the space to talk about their emotions, and could talk about their emotion better within groups with the help of the toolkit.
    \item Students will have better sense of belonging/feel supported in class.
    % \item Students will emotionally feel better after sharing with others.
    \item Students will understand struggles are normal in computing education, and everyone has them.
    % \item Students are able to learn from each other's strategies, coming up with their own strategies and use some strategies when doing their work.
\end{itemize}

Based on these outcomes, the authors met regularly over a period of a few weeks to come up with potential creative tasks and prompts for the cards. These were captured on Miro\footnote{\url{https://miro.com/}}. The final brainstorm from these discussions is shown in Figure \ref{fig:brainstorm}. It shows how we separated the reflective prompts on the cards into three categories: 

    \paragraph{\textcolor[HTML]{3fceec}{\textbf{\texttt{To prompt feelings:}}}} These cards (pictured in blue) focus on encouraging students to identify their current feelings. This included a variety of ways to express feelings such as to create clay sculptures \cite{Clay2026} and to annotate body maps \cite{bodymaps}. \textbf{Link to Reflection Literature:} \citet{atkins1993reflection} and \citet{boud1985promoting} show that recognition of one's feelings serves as an initial catalyst for self-reflection.
    
    \paragraph{\textcolor[HTML]{FF1493}{\textbf{\texttt{To programming skills:}}}} These cards (pictured in pink) relate more directly to the programming process, giving students new perspectives on their code to reflect on. This included to complete a diagram of the knowledge they already know and need to learn \cite{Ogle1986KWL} and to talk through their code with a rubber duck \cite{Hunt1999Pragmatic}. \textbf{Link to Reflection Literature:} \citet{bentvelzen2022revisiting} describes how enabling new and different perspectives can help to spark reflection.
    
    \paragraph{\textcolor[HTML]{50c878}{\textbf{\texttt{Break activities:}}}}  These cards (pictured in green) include novel activities to encourage self-regulation. Whilst not directly prompting reflection per se, the cards give opportunity for students to step away from their activity and take time off. Some examples are engaging in meditation and to walk backwards through a room.
    \textbf{Link to Reflection Literature:} \citet{fleck_reflecting_2010} suggests that stepping back can help to create the conditions for reflection; even if the cards don’t directly prompt reflection, they may still support it indirectly by providing space to reflect. \newline{} 

After multiple rounds of brainstorming and discussion, we settled on a final list of twenty-one prompts for the cards. Using InDesign\footnote{\url{https://www.adobe.com/uk/products/indesign.html}}, these prompts were designed as small, hand-held cards. Cards are coloured according to the category they belong to. The overall aesthetics of the cards are meant to convey playfulness and create a contrast to the (aesthetics of the) programming environments that students use.

\section{Evaluation} \label{sec:method}
A mixed-methods study took place over two consecutive teaching weeks at the end of a CS1 course. Each week, the session was identical, but with different students. Participants received an information sheet (read aloud in-class) and provided written informed consent. Participation in teaching activities was unaffected for non-consenting students (we removed the data from these students in the class). The study was approved by the University of the Arts London ethics committee. 

\subsection{Participants}\label{sec:participants}
Participants were first-year students on the Intro to CS module at the University of the Arts London, shared between BSc (Hons) Computer Science and BSc (Hons) AI and Data Science degrees. 29 students participated in total. 

\subsection{Procedure \& Data Collection}
The procedure for each session was as follows: 
\begin{enumerate}
    \item Participants sign consent forms and are given anonymous ID numbers.
    \item Participants completed a pre-survey. Participants' background information on when they started programming and their familiarity with different tools were collected. We also collected age, gender, and where participants spent their formative years. 
    
    The PANAS \cite{Watson1988PANAS} questionnaire was also taken at this point to collect a baseline measure of how participants were feeling before the study start. However, in this paper, we focus on the qualitative insights.
    
    \item Participants were given 45 mins to start on a programming assignment, using a brief from previous years. The brief guides participants through creating a P5js\footnote{\url{https://p5js.org/}} project making ASCII art --- taking inspiration from Dwarf Fortress \cite{Adams2006Dwarf} and the Terraforms by Mathcastles collection \cite{Mathcastles2021Terraforms}.  
    \item Participants completed the PANAS questionnaire again.
    \item Participants were asked to pick from the \textcolor[HTML]{3fceec}{\textbf{\texttt{feelings}}} and \textcolor[HTML]{FF1493}{\textbf{\texttt{programming skills}}} cards for 15 mins.
    \item Next, a focus group was conducted in-class by two researchers different to the main instructor. This helped to reduce bias in students who might confirm what they expected their lecturer wanted to hear. The focus group was recorded using the classroom's Panopto\footnote{\url{www.panopto.com/}} system and microphones, capturing the discussion without being obtrusive to the teaching and learning context. The focus group was open-ended with the following prompting questions: What did you do?; How did it make you feel?; and How did it influence your programming practice?. 
    \item Participants were asked to pick from the \texttt{\textbf{\textcolor[HTML]{50c878}{break activities}}} cards for 15 mins.
    \item The focus group in Step 6 was repeated.
    \item A post survey repeats the PANAS questionnaire following the design card activities and opportunity is given for any final comments.
\end{enumerate}

\noindent A week after the taught sessions participants were invited to take part in a 1:1 interview to more deeply understand their opinions and perceptions of the card activities.

\section{Initial Insights}\label{sec:findings}
Below we present our initial observations from our early impressions of the focus group and interview recordings. We plan a more detailed thematic analysis \cite{braun_reflecting_2019} at a later date.

\subsection{Observation: Stepping away to take a break}
Some students found that the \textcolor[HTML]{50c878}{\textbf{\texttt{break activity}}} cards helped them to step away from the initial pressure to code. One student described the activity as ``allow[ing] me to step out and take some time off to really like take a break'', while another explained that ``if I leave I can have a think and go back to it[and then] I sometimes remember [what to do]''. Rather than directly helping to solve issues in their coding, students instead described the cards as helping them relax. One student reflected that ``it helped me to kind of relax a bit to, kind of, think about something else, cause I was stuck for like 15 mins at the end of the assignment''.

\subsection{Observation: Changing location to socialise}
Several students described how physically moving away from their desks created opportunities for shared reflection. Students explained that while walking they were ``just discussing about the task'', and that ``recapping on the walk helped to understand the task more''. Others described how changing location supported thinking through confusion and uncertainty together; one student with their peers ``went outside by the kitchen and [discussed] how to actually approach the task, which [they] found quite difficult''. 

\subsection{Observation: Some tasks are too weird and disrupt coding flow}
Activities which initially seemed playful or unusual sparked opportunities for reflection. For example, one student described how they ``took a 5 min walk with a duck, because [they] wanted to do something with it, whilst thinking about the [assignment] confusion''. 

However, other students, whilst appreciating the playful nature of the cards, felt that the tasks were too disconnected from their usual programming practice. One student noted that such tasks ``just reminded me that I need to get back in the mindset of coding'', whilst another commented that ``after 45 mins if I go and do an exercise about ducks, I feel there is a time and a place, but I don't feel it should be designated''. Furthermore, these students -- who tended to have stronger programming skills than those who enjoyed the playful nature of the cards -- responded more positively to self-regulation activities which they perceived as more natural or familiar such as taking breaks or walking. One student explained that they were ``not a big fan of doing random things like jumping jacks etc, but yeah I like the idea of taking a break''.

\subsection{Observation: Breaking things down}
Many students found the \textcolor[HTML]{FF1493}{\textbf{\texttt{programming skills}}} cards helpful. One student explained that the activities ``would make me want to... break the things down to an even more step by step''. Others connected this to managing feelings of overload and time pressure. For example, one student described how they needed to ``break my project into smaller manageable stuff as my brain just overloads, and there was a short time limit and it freaked me out abit''. Students also found value in prompts that externalised and organised their thinking processes. For example, one student reflected on the K-W-L \cite{Ogle1986KWL} task: ``the three columns... help organise my brain''.

\subsection{Observation: Making helped express emotion if not time consuming}
\begin{figure}[h!]
    \centering
    \includegraphics[width=0.7\linewidth]{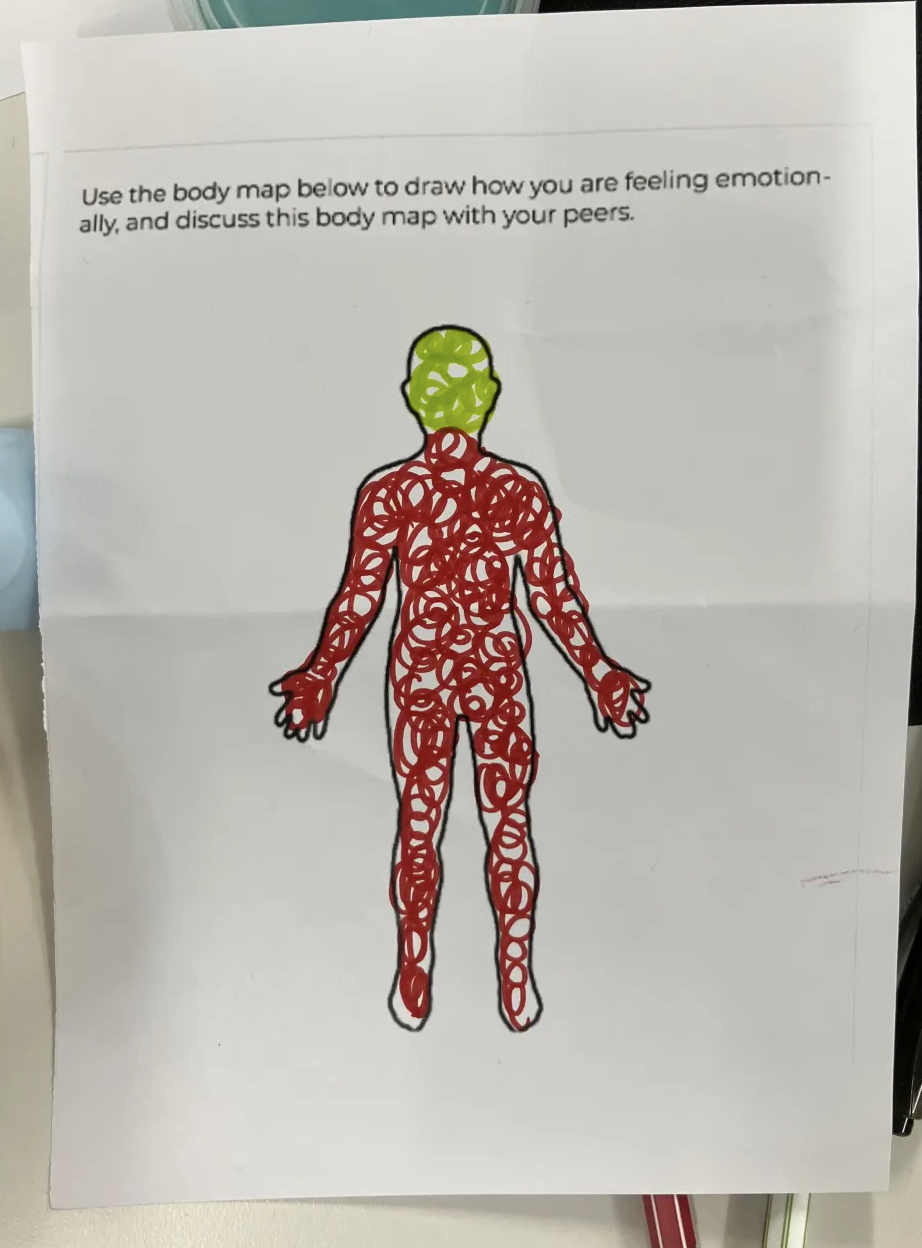}
    \caption{Example body map used by a student to express their feelings whilst coding.}
    \vspace*{0.2cm}
    \label{fig:map}
\end{figure}

\begin{figure}[h!]
    \centering
    \includegraphics[width=0.55\linewidth]{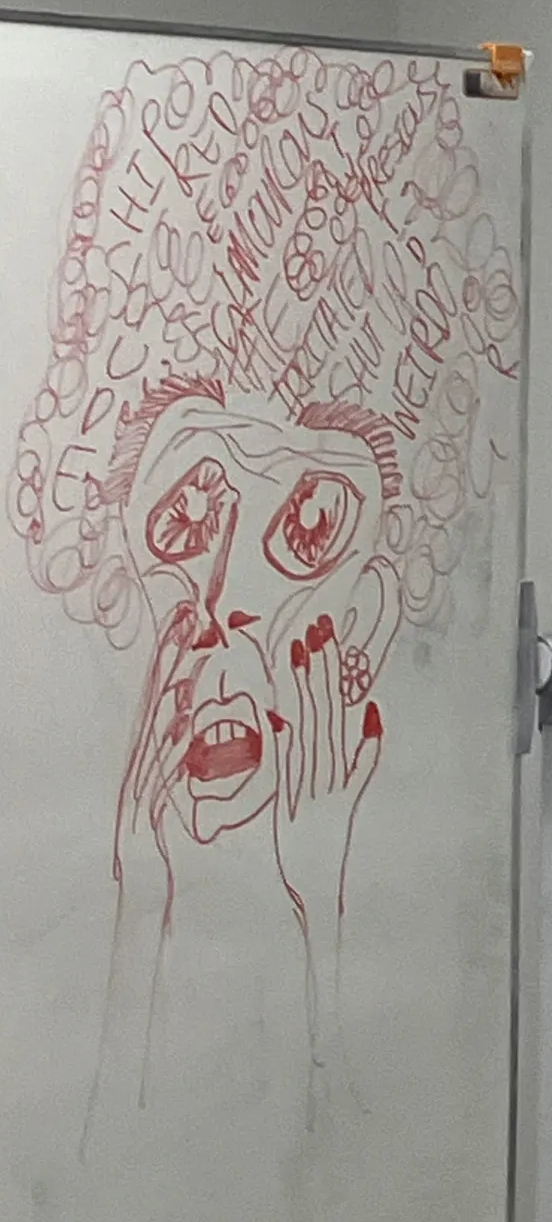}
    \caption{Example drawing used to represent a students feelings whilst coding. }
    \label{fig:scrib}
\end{figure}

Some students used the more creative and material-focused prompts as ways to externalise and communicate emotions associated with programming. One student explained that ``using the map helped to express my confusion'' (see Figure \ref{fig:map}). Another drew a picture (see Figure \ref{fig:scrib}) with ``trembling hands'' and hair made of ``words that[...] reflect all like depressing stuff but the right hand side is quite positive''. Although they found ``trying to get the drawing right'' somewhat stressful. Others avoided the comic tasks -- they thought it would take too much time. 

\section{Summary}\label{sec:discussion}
This paper examined the question: what if we ask students to use creative task cards in peer-to-peer discussions which prompt reflection on emotional aspects of programming; how does this support students’ self-efficacy and self-regulation?

From our preliminary observations, we suggest that students self-efficacy and self-regulation is supported when:
\begin{itemize}
    \item Students can recognise frustration as a shared experiences in programming, particularly through peer discussions during break activities and walks.
    
    \item Students can externalise and communicate emotions associated with programming through creative reflective tasks, such as body maps and drawings.
    
    \item Students use reflective prompts that help break programming tasks into smaller and more manageable steps, reducing feelings of cognitive overwhelm.
    
    \item Students are provided with opportunities to pause and emotionally reset during programming activities, helping them re-engage with tasks after becoming stuck or frustrated. 
\end{itemize}

We look forward to showing more examples and discussing our insights and design process in more detail at the RiCE workshop.

%=========================================================

% Uncomment 
% \begin{acks}

% \end{acks}

\bibliographystyle{ACM-Reference-Format}
\bibliography{references}

% \section*{Appendix}

% \appendix  %<- You submit an appendix, but do not need to link to it for the assessment! :)

\end{document}